\documentclass[aps, prd, twocolumn, superscriptaddress, nofootinbib]{revtex4}
\usepackage{graphicx}
\usepackage{dcolumn}
\usepackage{amssymb,amsmath,amsthm,mathrsfs}

\begin{document}

\newcommand{\Eq}[1]{Eq. \ref{eqn:#1}}
\newcommand{\Fig}[1]{Fig. \ref{fig:#1}}
\newcommand{\Sec}[1]{Sec. \ref{sec:#1}}

\newcommand{\PHI}{\phi}
\newcommand{\vect}[1]{\mathbf{#1}}
\newcommand{\Del}{\nabla}
\newcommand{\unit}[1]{\mathrm{#1}}
\newcommand{\x}{\vect{x}}
\newcommand{\ScS}{\scriptstyle}
\newcommand{\ScScS}{\scriptscriptstyle}
\newcommand{\xplus}[1]{\vect{x}\!\ScScS{+}\!\ScS\vect{#1}}
\newcommand{\xminus}[1]{\vect{x}\!\ScScS{-}\!\ScS\vect{#1}}
\newcommand{\diff}{\mathrm{d}}

\newcommand{\be}{\begin{equation}}
\newcommand{\ee}{\end{equation}}
\newcommand{\bea}{\begin{eqnarray}}
\newcommand{\eea}{\end{eqnarray}}
\newcommand{\vu}{{\mathbf u}}
\newcommand{\ve}{{\mathbf e}}
\newcommand{\vk}{{\mathbf k}}
\newcommand{\vx}{{\mathbf x}}
\newcommand{\vy}{{\mathbf y}}

\newcommand{\uden}{\underset{\widetilde{}}}
\newcommand{\den}{\overset{\widetilde{}}}
\newcommand{\denep}{\underset{\widetilde{}}{\epsilon}}

\newcommand{\nn}{\nonumber \\}
\newcommand{\dd}{\diff}
\newcommand{\fr}{\frac}
\newcommand{\del}{\partial}
\newcommand{\eps}{\epsilon}
\newcommand\CS{\mathcal{C}}

\def\be{\begin{equation}}
\def\ee{\end{equation}}
\def\ben{\begin{equation*}}
\def\een{\end{equation*}}
\def\bea{\begin{eqnarray}}
\def\eea{\end{eqnarray}}
\def\bal{\begin{align}}
\def\eal{\end{align}}


\title{New ground state for quantum gravity}

\newcommand{\addressImperial}{Theoretical Physics, Blackett Laboratory, Imperial College, London, SW7 2BZ, United Kingdom}
\author{Jo\~{a}o Magueijo}
\author{Laura Bethke}

\affiliation{\addressImperial}

\date{\today}


\begin{abstract}
In this paper we conjecture the existence of a new ``ground'' state 
in quantum gravity, supplying a wave function for the inflationary
Universe. We present its explicit perturbative expression
in the connection representation, exhibiting the associated inner product.
The state is chiral, dependent on the Immirzi parameter, and is the 
vacuum of a second quantized theory of graviton particles. 
We identify the physical and unphysical Hilbert sub-spaces.
We then contrast this state with the perturbed Kodama state and 
explain why the latter can never describe gravitons in a de Sitter background.
Instead, it describes self-dual excitations,
which are composites of the positive frequencies of the right-handed 
graviton and the negative frequencies of the left-handed graviton.
These excitations are shown to be unphysical under the inner product 
we have identified. 
Our rejection of the Kodama state has a moral tale to it: 
the semi-classical limit of quantum gravity can be the wrong
path for making contact with reality (which may
sometimes be perturbative but nonetheless fully quantum). Our results
point towards a non-perturbative extension, and we present some 
conjectures on the nature of this hypothetical state.
\end{abstract}


\maketitle


\section{Introduction}

Although the modern canonical approach to quantum 
gravity~\cite{ashbook,rovbook,thbook} has 
in many ways been a formal success, it has not always made easy contact
with the real world (see, 
however,~\cite{recentgravitons,rovrecent,lqc,lqcnew,dip}). 
This is often blamed not on the formalism, but on the difficulty in 
finding solutions representing something akin to the reality 
we observe, possibly corrected by new effects still beyond the reach of 
experiment. The matter is closely related to the identification of the 
ground state (or ``base'' state, a terminology
intended to skirt the issue of energy) around which small
excitations would form the reality we probe. In some approximation
the theory should produce solutions representing 
classical space-times satisfying Einstein's equations,
subject to small quantum fluctuations 
or semi-classical corrections. In a cosmological setting 
the theory should supply a wave-function for the early Universe, 
for example in a de Sitter or inflationary primordial phase. This
wave function should encode predictions for the vacuum 
quantum fluctuations that seed the structure of our Universe
or make up a cosmic gravitational wave background. 
Ultimately quantum gravity should corroborate---or contradict---the 
textbook treatment~\cite{lid,muk}, based on second quantized 
effective field theory and the Bunch-Davis vacuum.

One of the earliest proposals for a ``base state'' describing de Sitter
space-time was the Kodama state~\cite{kodama}, sometimes referred to as
the Chern-Simons wave function. This state solves the Hamiltonian
constraint in a diagonal representation for the  Ashtekar connection,
with ``$EEF$'' factor ordering, cosmological constant $\Lambda\neq 0$,
and Immirzi parameter $\gamma=\pm i$. 
As such, the Kodama state should be the ideal candidate for 
the wave function of the inflationary Universe, expressed in terms of the
Ashtekar self-dual (or anti-self-dual) connection, instead of the 
more traditional metric representation. Furthermore, the
Chern-Simons functional is the Hamilton-Jacobi function of the theory, 
so that  the Kodama state is a semi-classical or WKB solution even 
if the space-time is only approximately de Sitter. 
It should therefore provide the vacuum 
for inflationary fluctuations, placing existing heuristic calculations 
on safe grounds. 

Regrettably, in spite of earlier
claims to that effect~\cite{leereview,leelaur}, it has proved elusive
and even self-contradictory~\cite{wittenym,leelaur} 
finding a second quantized
theory of linearized gravitons inside the perturbed Kodama state.
Tensor modes around de Sitter space-time simply 
do not fit into the perturbed Kodama state.
In this paper we explain the reasons for this failure and exhibit
the true perturbative state representing gravitons
in the same representation used for the Kodama state.
We will do this for the same $\gamma$ and ordering of the 
Hamiltonian constraint as that implicit in the construction of the Kodama state,
but also more generally.

In recent work~\cite{prl,paper,gengam} we derived explicit perturbative
expressions for the Fock vacuum and its graviton states, using perturbative
expansions in Ashtekar gravity. These papers built on earlier 
research~\cite{ash0,gravitons}, and lead to the conclusion that, even though
the graviton particle spectrum is identical for right and left gravitons (in
contrast with the findings of~\cite{wittenym}), their vacuum energy and
fluctuations have a chiral signature (dependent on $\gamma$
and factor ordering issues). The observational implications of this
conclusion are striking, as
discussed in~\cite{TBpapers}. The first task of the present paper is to
express these results in the same representation as that used
for the Kodama state, i.e. a holomorphic connection representation.
A number of technical issues, not evident in the Bargmann representation 
used in~\cite{prl,paper}, are discussed in Section~\ref{connect}, following
which explicit forms for the wave functions are found (Section~\ref{ground}).
Their norms are evaluated in Section~\ref{norms}, selecting the physical and 
unphysical sub-spaces.

Direct comparison of these states with the perturbed Kodama state reveals
blatant antagonism. In Section~\ref{kodama} we identify its origin.
We show that even before issues related to the inner product are considered
(normalizability, the physical sub-space, etc), it is obvious that
{\it the perturbed Kodama state is not an eigenstate of the perturbative 
Hamiltonian}, and so can never represent gravitons or their vacuum 
in a second quantized theory of tensor perturbations
around de Sitter space-time. 
This arises from the simple algebraic fact
that the perturbed self-dual operator does not factor out in
the perturbed Hamiltonian, even though it does so non-perturbatively.
This is far from an oddity and results from the ``braided'' fashion in which 
perturbative expansions  contribute to a non-linear expression at a given
order, as spelled out in Section~\ref{simple}.

Our conclusion is far from surprising. 
The perturbed Kodama state represents solutions to the 
perturbed self-dual operator, i.e. self-dual excitations. 
These have long been known~\cite{ash0} to combine the 
positive frequency of the right-handed graviton and the negative frequency
of the left-handed anti-graviton (see~\cite{paper} for further discussion). 
As in the age-old adage, two halves don't make a whole: not only are 
these composites of half-particles not gravitons, but they are
unphysical, as can be checked by evaluating their norm
using the same criteria employed to select physical graviton states.
This conclusion is also true for the Chern-Simons state in
Yang-Mills theories, although the reason there is more subtle
in this case.

Having presented a solution for the base state and discredited its 
competition we close this paper by speculating on 
possible non-perturbative generalizations for our construction
(Section~\ref{specs}).

\section{The connection representation of gravitons}\label{connect}
In~\cite{prl,paper} we obtained an expression for the perturbative 
Hamiltonian in the Ashtekar formalism by expanding the Hamiltonian constraint
to second order, retaining only terms quadratic in the first order expansions.
We refer the reader to~\cite{paper} for notation, definitions and 
all the subtleties. The upshot is the identification of a 
set of particle creation and annihilation graviton operators, 
${\cal G}_{r{\cal P}}^\dagger$ and ${\cal G}_{r{\cal P}}$,
combining metric and connection variables. The theory contains 
gravitons and anti-gravitons before the reality conditions are imposed
and, due to their second class nature, these conditions are imposed via
the choice of the inner product with which the Hilbert space is endowed.
The inner product then flags half the particles as unphysical, restoring the 
correct number of physical 
degrees of freedom in the theory. In~\cite{prl,paper} we
performed this exercise in representations which diagonalize
the creation operator---the so-called Bargmann 
representation~\cite{barg0,barg1,bargrev}. For each value of $\gamma$
we obtain a different representation. Here we recast these results in 
the connection representation, mimicking the format in which the Kodama
state is expressed.

It is not immediately obvious that a holomorphic representation
diagonalizing the connection exists at all. 
Indeed, in a strict mathematical sense, 
such representations do not exist for $\gamma=\pm i$, as we shall
see. Yet, wave functions in the Ashtekar approach are usually expressed
as functions of the connection or its holonomies. 
In this Section we show how to get
as close as possible to a holomorphic representation diagonalizing
the connection, adapting the usual construction.

There are different terminologies in use, so to be 
completely clear let us
define as {\it holomorphic} a representation in terms of 
complex functions $\Psi(z)$ which depend on a
set of complex variables $z$ but not their complex conjugates
${\bar z}$ ($z$ and ${\bar z}$ are to be seen as independent
variables when differentiating or integrating). If in addition
these functions are analytic functions over the whole complex domain 
(i.e. they can be expressed as a power series over the whole domain)
we call them {\it entire} functions. This distinction will be essential 
later.

Using this terminology, the functions forming the basis for 
the Bargmann representation are entire functions. The problems encountered
searching for the  dual of this representation
(diagonalizing the annihilation operator, instead
of the creation operator) whilst keeping the functions entire have
been discussed, for example, in~\cite{ribeiro,ribeiro1}. 
The matter is relevant for discussing wave-functions when $\gamma=\pm i$, 
(and more generally for the definition of a delta
function in the complex domain), and it will be discussed in  
Section~\ref{groundsd}. More generally, even for $\gamma\neq\pm i$,
finding a holomorphic representation (in the sense defined above)
diagonalizing the connection 
proves  problematic. Since our proposal will not be the 
obvious one, we start by spelling 
out the problem before providing the alternative.

\subsection{The apparent obstruction}\label{connect-wrong}
As in~\cite{paper} (where we focused
on a purely imaginary $\gamma$) we expand connection and 
metric perturbations into Fourier modes ${\tilde a}_{rp}(\vk)$ 
and ${\tilde e}_{rp}(\vk)$, where
$r=\pm$ represents right and left polarizations, and $p=\pm $ 
indexes the graviton
and anti-graviton modes (needed to expand variables which, a priori, 
are complex).
We then find, from their Poisson bracket, that upon quantization the 
operators to which they are promoted should satisfy:
\be\label{fixedcrs} [{\widehat a}_{rp}(\vk),{\widehat
e}_{sq}^\dagger(\vk ')] =-i\gamma p
\frac{l_P^2}{2}\delta_{rs}\delta_{p{\bar q}} \delta(\vk-\vk ')\; ,
\ee
with all other commutators set to zero. We want to represent 
this algebra. Since ${\widehat a}_{rp}$ commutes with
${\widehat a}^\dagger_{rp}$, we might be tempted to think that if 
we diagonalize one of them 
we diagonalize the other, a statement which relies on the crucial 
assumption that {\it they act on the same space of functions}. 
This is a trap leading to serious problems, which we spell out here
in order to  motivate the alternative construction presented
in Section~\ref{hol-solution}.

Let us then accept that ${\widehat a}_{rp}$ and ${\widehat a}^\dagger_{rp}$
act on the same space of functions. Since they commute, they can be
simultaneously diagonalized, so that ${\widehat a}\Psi=\lambda\Psi$ 
and ${\widehat a}^\dagger
\Psi=\mu\Psi$. From $\langle \Psi|{\widehat a}^\dagger|\Phi\rangle=
{\overline{\langle \Phi|{\widehat a}|\Psi\rangle}}$ we can conclude $\mu=
\overline\lambda$, i.e. the eigenvalues of $a_{rp}^\dagger$ must be 
complex conjugate to those of $a_{rp}$. 
It is therefore inevitable that functions $\Phi$ must mix variables 
and their conjugates, 
i.e. $\Phi=\Phi(a_{rp},{\bar a}_{rp})$. The representation
can no longer be holomorphic, and we have instead:
\be\label{arp}
{\widehat a}_{rp}\Phi(a_{rp},\overline{a}_{rp})={a}_{rp}\Phi(a_{rp},
\overline{a}_{rp})
\ee
\be\label{arpdagger}
{\widehat a}^\dagger _{rp}\Phi(a_{rp},\overline{a}_{rp})=
\overline{a}_{rp}\Phi(a_{rp},\overline{a}_{rp})
\ee
with commutation relations (\ref{fixedcrs}) implying:
\be\label{erp}
{\widehat e}^\dagger_{rp} \Phi(a_{rp},\overline{a}_{rp})
=-i\gamma p\frac{l_P^2}{2}\frac{\delta}{\delta
a_{r{\bar p}}}\Phi(a_{rp},\overline{a}_{rp})
\ee
\be\label{erpdagger}
{\widehat e}_{rp} \Phi(a_{rp},\overline{a}_{rp})
=i\gamma p\frac{l_P^2}{2}\frac{\delta}{\delta
{\overline a}_{r{\bar p}}}\Phi(a_{rp},\overline{a}_{rp})\; .
\ee
The fact that the representation is no longer holomorphic leads to the 
breakdown of several results, namely the standard derivation of 
the inner product. Indeed, the anzatz for the inner product 
would now have to be:
\be\label{ansatz}{\langle
\Phi_1 | \Phi_2\rangle}=\int d a \, d {\bar {a}}\,  e^{\mu(a,{\bar a})}
{\bar \Phi_1}({\bar a}, a) \Phi_2 (a, {\bar a})\; ,\ee
(note the dependence on $a$ and $\bar a$ of both functions).
The reality conditions are:
\bea{\widehat a}_{r+} +
{\widehat a}_{r-} &=& 2 r k
{\widehat e}_{r+}\label{real1}\\
{\widehat a}^\dagger_{r+} + {\widehat a}^\dagger_{r-} &=&
2 r k {\widehat e}^\dagger_{r-}\; ,\label{real2}\\
{\widehat e}_{r+}&=&{\widehat e}_{r-}\label{real3}
\eea 
and these are formally valid only when sandwiched between a generic 
bra and a ket. In the standard treatment~\cite{tate,pulbook,gravitons} 
this fully fixes $\mu$, but in the derivation one needs to perform an 
integration by parts which assumes that the functions are holomorphic.
With non-holomorphic functions this can no longer be done
(since both $a$ and $\bar a$ appear in each of $\bar \Phi_1$ and 
$\Phi_2$). It is straightforward to show that erroneously
neglecting this detail leads to contradictory
conditions regarding the sign of $\mu$.

\subsection{The resolution}\label{hol-solution}
It is possible to define a holomorphic representation in
connection space, but as the previous sub-Section showed, one
must drop the assumption that ${\widehat a}_{rp}$ and ${\widehat a}_{rp}^\dagger$ 
act on the same space of functions. In fact this feature is suggested
by the formalism. By direct inspection (see~\cite{paper}) we can check
that the only operators that appear in the Hamiltonian
(on- and off-shell) are ${\widehat a}_{r+}$, ${\widehat a}_{r-}^\dagger$, 
${\widehat e}_{r+}$ and ${\widehat e}_{r-}^\dagger$,
a feature which propagates into the definition of creation and annihilation
operators, ${\cal G}^\dagger_{r{\cal P}}$ and ${\cal G}_{r{\cal P}}$.
This suggests restricting our ket functions to be functions of
of ${a}_{r+}$ and  ${\overline a}_{r-}$ only, noting that 
${\widehat e}_{r+}$ and ${\widehat e}_{r-}^\dagger$ will naturally act
on them, given commutation relations (\ref{fixedcrs}). The operators
${\widehat a}_{r+}$, ${\widehat a}_{r-}^\dagger$, 
${\widehat e}_{r+}$ and ${\widehat e}_{r-}^\dagger$,
therefore act {\it on the right} upon functions 
$\Phi=\Phi({a}_{r+}, {\overline a}_{r-})$. The dual space of 
functions (``bra'' functions) will then be functions of the conjugate 
variables, ${\overline \Phi}= {\overline \Phi}
({\overline a}_{r+}, {a}_{r-})$, with the conjugate operators, 
${\widehat a}^\dagger_{r+}$, ${\widehat a}_{r-}$, 
${\widehat e}^\dagger_{r+}$ and ${\widehat e}_{r-}$,
acting {\it on the left} upon them.

Spelling this out\footnote{
In this paper we 
adopt the convention
$\Psi(\lambda)={\langle \lambda |\Psi \rangle}$
for ``ket'' functions, and 
${\bar \Psi}({\bar \lambda})={\langle \Psi|\lambda \rangle}$
for the dual ``bra'' functions. Other conventions are possible, 
e.g.~\cite{ribeiro}.},  we postulate a Hilbert space of
(ket) wavefunctions holomorphic in $a_{r+}$ and ${\bar a}_{r-}$:
\be
\Phi(a_{r+},{\bar a}_{r-})
={\langle a_{rp}|\Phi\rangle}
\ee
and a dual space of (bra) functions holomorphic in ${\bar a}_{r+}$
and $a_{r-}$:
\be
{\bar \Phi}({\bar a}_{r+}, a_{r-})
={\langle \Phi |a_{rp} \rangle}\; .
\ee
Instead of (\ref{arp})-(\ref{erpdagger}),
we have:
\bea
{\widehat a}_{r+}\Phi(a_{r+},{\overline a}_{r-})&=&{a}_{r+}\Phi(a_{r+},{\overline a}_{r-})\label{newarp}\\
{\widehat a}^\dagger_{r-}\Phi(a_{r+},{\overline a}_{r-})&=&{\bar a}_{r-}\Phi(a_{r+},{\overline a}_{r-})\label{newarmdag}\\
{\widehat e}_{r+}\Phi(a_{r+},{\overline a}_{r-})&=&i\gamma \frac{l_P^2}{2}\frac{\delta}{\delta
{\bar a}_{r-}}\Phi(a_{r+},{\overline a}_{r-})\label{newerp}\\
{\widehat e}^\dagger _{r-}\Phi(a_{r+},{\overline a}_{r-})&=&i\gamma
\frac{l_P^2}{2}\frac{\delta}{\delta a_{r+}}\Phi(a_{r+},{\overline a}_{r-}) 
\label{newermdag}
\eea for operators
acting on the right upon ket functions $\Phi=\Phi(a_{r+},{\bar
a}_{r-})$. Reciprocally, we have:
\bea
{\bar \Phi}({\overline a}_{r+},a_{r-}){\widehat a}^\dagger _{r+}&=&{\bar \Phi}({\overline a}_{r+},a_{r-}){\bar a}_{r+}\\
{\bar \Phi}({\overline a}_{r+},a_{r-}){\widehat a}_{r-}&=&{\bar \Phi}({\overline a}_{r+},a_{r-})a_{r-}\label{ar-}\\
{\bar \Phi}({\overline a}_{r+},a_{r-}){\widehat e}^\dagger _{r+}&=&{\bar \Phi}({\overline a}_{r+},a_{r-}){\left(
i\gamma \frac{l_P^2}{2}\frac{\overleftarrow\delta}{\delta
{a}_{r-}}\right)}\label{newerpdag}\\
{\bar \Phi}({\overline a}_{r+},a_{r-}){\widehat e}_{r-}&=&{\bar \Phi}({\overline a}_{r+},a_{r-}){\left( i\gamma
\frac{l_P^2}{2}\frac{\overleftarrow\delta}{\delta {\bar
a}_{r+}}\right)} \label{newerm}
\eea 
for operators acting on the left, upon dual functions
$\bar \Phi=\bar \Phi({\bar a}_{r+}, a_{r-})$. 
In both cases the algebra (\ref{fixedcrs}) is realized,
but notice that there is a minus sign in the
last two formulae, (\ref{newerpdag}) and (\ref{newerm}), 
with respect to (\ref{erp}) and (\ref{erpdagger}),
resulting from the fact that the operators act on the left, not 
on the right.
With this prescription operators never map a holomorphic function
into a non-holomorphic function,  and since operators and their 
conjugates never act upon the same space, we evade the ``trap'' 
highlighted in the previous Section, which forced the functions to be
non-holomorphic. 

Within this set up we can now determine the inner product from the reality
conditions, following the usual procedure. In contrast to (\ref{ansatz})
we consider ansatz: \be\label{ansatz1}{\langle \Phi_1 |
\Phi_2\rangle}=\int d a \, d {\bar {a}}\,  e^{\mu(a,{\bar a})} {\bar
\Phi_1}({\bar a}_{r+}, a_{r-}) \Phi_2 (a_{r+}, {\bar a}_{r-})
\nonumber\ee
where the integration measure mixes variables and their
conjugates, but the wavefunctions and their duals do not. The
reality conditions  are still (\ref{real1}), (\ref{real2}) and (\ref{real3}).
These mix right and left acting
operators, however the reality conditions are {\it not} to be seen
as operator conditions, but as conditions upon the inner product.
Therefore they should always be sandwiched between a bra and a ket
and only make sense in that context. Condition
${\widehat e}_{r+}={\widehat e}_{r-}$, for example, does not make
sense as an operator condition
(it's like imposing an identity between different types of
objects), but ${\langle \Phi|{\widehat e}_{r+}|\Psi\rangle}
={\langle \Phi|{\widehat e}_{r-}|\Psi\rangle}$ does.

Following the standard argument, conditions (\ref{real1}) and (\ref{real2})
lead to:
\bea a_{r+} +
a_{r-}&=&-i\gamma l_P^2 kr\frac{\delta \mu}{\delta {\bar a}_{r-}}\\
{\bar a}_{r+} + {\bar a}_{r-}&=&-i\gamma l_P^2 kr\frac{\delta
\mu}{\delta {a}_{r+}} \eea with solution: \be\label{solmu}
\mu=
\int d^3k \sum_r
\frac{-1}{i\gamma l_P^2 kr}(a_{r+}+a_{r-})({\bar a}_{r+}+{\bar
a}_{r-}) \; .\ee 
In contrast to the apparent sign contradiction mentioned at the end of 
Section~\ref{connect-wrong}, identities 
${\widehat e}_{r+}={\widehat e}_{r-}$ and ${\widehat
e}_{r+}^\dagger={\widehat e}_{r-}^\dagger$ now merely
signify: \bea
\frac{\delta \mu}{\delta {\bar a}_{r-}}&=& \frac{\delta
\mu}{\delta {\bar a}_{r+}}\\
\frac{\delta \mu}{\delta {a}_{r-}}&=&\frac{\delta \mu}{\delta
{a}_{r+}} \eea
satisfied by solution (\ref{solmu}). 
We note that $\mu$ is real, as it should be. 

\subsection{Result for a general complex $\gamma$}
The results shown thus far (as well as those in~\cite{paper}) are valid
for a purely imaginary $\gamma$. For a generally complex $\gamma$ 
(investigated in~\cite{gengam}), the reality conditions 
(\ref{real1})-(\ref{real2})-(\ref{real3})
should be replaced by:
\bea  i\gamma^\ast{\widehat a}_{r+}- i\gamma{\widehat a}_{r-}&=& 
2 r k \gamma_I {\widehat e}_{r+} \label{gengamreal1}\\
-i\gamma {\widehat a}^\dagger_{r+} + i\gamma^
\ast{\widehat a}^\dagger_{r-} &=& 
2 r k \gamma_I {\widehat e}^\dagger_{r-}\\ 
{\widehat e}_{r+}&=&{\widehat e}_{r-} \; .
\eea
Additionally, the commutation relations (\ref{fixedcrs}) become:
\be [{\widehat a}_{rp}(\vk),{\widehat
e}_{sq}^\dagger(\vk ')] =-i(\gamma_R+pi\gamma_I)
\frac{l_P^2}{2}\delta_{rs}\delta_{p{\bar q}} \delta(\vk-\vk ')\; .
\ee
In a holomorphic representation that diagonalises the connection, as above,
this leaves equations (\ref{newarp}) to (\ref{ar-}) unmodified, 
but (\ref{newerpdag}) and (\ref{newerm}) change to:
\bea {\bar \Phi}{\widehat e}^\dagger _{r+}&=&{\bar \Phi}{\left(
-i\gamma^\ast \frac{l_P^2}{2}\frac{\overleftarrow\delta}{\delta
{a}_{r-}}\right)}\\
{\bar \Phi}{\widehat e}_{r-}&=&{\bar \Phi}{\left(-i\gamma^\ast
\frac{l_P^2}{2}\frac{\overleftarrow\delta}{\delta {\bar
a}_{r+}}\right)} \; .\eea
Using the standard approach, we can find 
functional differential equations for the measure:
\bea i\gamma^\ast a_{r+} -i\gamma
a_{r-}&=&-i\gamma l_P^2 kr\gamma_I\frac{\delta \mu}{\delta {\bar a}_{r-}}
\label{eqmu1}\\
-i\gamma{\bar a}_{r+} + i\gamma^\ast {\bar a}_{r-}&=&-i\gamma l_P^2 kr\gamma_I\frac{\delta
\mu}{\delta {a}_{r+}} \eea 
with the reality of the metric implying:
\be -i\gamma \frac{\delta \mu}{\delta {\bar a}_{r-}} = 
i\gamma^\ast \frac{\delta \mu}{\delta {\bar a}_{r+}} \; . \ee
The measure for a generally complex $\gamma$ is therefore:
\bea
\mu &=&\int d^3k \sum_r \frac{1}{l_P^2kr\gamma_I}\nonumber \\
&&\bigg[a_{r+}{\bar a}_{r+} + a_{r-}{\bar a}_{r-}  
- \left(\frac{\gamma^\ast}{\gamma}a_{r+}{\bar a}_{r-} + 
\frac{\gamma}{\gamma^\ast}a_{r-}{\bar a}_{r+}\right) \bigg] \nonumber\eea
where we've assumed that $\gamma_I\neq 0$.

This reduces to (\ref{solmu}) in the limit $\gamma_R \rightarrow 0$ as required.
Note, however, that the limit $\gamma_I \rightarrow 0$ is ill-defined.
This is because in the case of a purely real $\gamma$ the representation
is no longer holomorphic. For $\gamma_I=0$, Eq.~(\ref{eqmu1})
becomes:
\be
a_{r+}=a_{r-}
\ee
which makes sense: since the theory is real we do not need the index $p$ .
This precludes the segregation of variables between kets ($a_{r+}$ and
${\bar a}_{r-}$) and bras (${\bar a}_{r+}$ and $a_{r-}$). If 
$\Phi(a_{r+},{\bar a}_{r-})$, then, dropping the redundant second
index, this means $\Phi(a_r,{\bar a}_r)$, i.e. the function is no longer
holomorphic. The procedure introduced
in this paper to find a connection holomorphic representation for 
the graviton states therefore does not
work for real $\gamma$.

\subsection{Absence of a metric holomorphic representation}
Another context where the prescription in Section~\ref{hol-solution}
breaks down, for the same reasons as in the previous subsection, 
is a representation diagonalizing the metric. This is very 
interesting: not only does the connection take precedence over the metric
in this formalism, but it appears that some results do not 
have counterparts expressed in terms of metric variables.

It may seem at first that
setting up a representation diagonalizing the metric is
a trivial extension of our method. In analogy with 
Section~\ref{hol-solution}, we should define kets such that
$\Phi=\Phi(e_{r+},\bar e_{r-})$ 
and bras such that $\bar\Phi=\bar\Phi(\bar e_{r+}, e_{r-})$ with the
following diagonal operators acting on the right:
\bea
{\widehat e}_{r+}\Phi &=& e_{r+}\Phi\\
{\widehat e}^\dagger_{r-}\Phi &=& \bar e_{r-}\Phi
\eea
and the remaining ones acting on the left:
\bea
\bar\Phi{\widehat e}^\dagger_{r+} &=& \bar\Phi \bar e_{r+}\\
\bar\Phi{\widehat e}_{r-} &=& \bar \Phi e_{r-}\; .
\eea
The commutation relations then lead to
\bea
{\widehat a}_{r+}\Phi&=&-i\gamma \frac{l_P^2}{2}\frac{\delta}{\delta
{\bar e}_{r-}}\Phi\\
{\widehat a}^\dagger _{r-}\Phi&=&-i\gamma
\frac{l_P^2}{2}\frac{\delta}{\delta e_{r+}}\Phi
\eea
for right-acting operators, and
\bea
{\bar \Phi}{\widehat a}^\dagger_{r+}&=&{\bar \Phi}
{\left(-
i\gamma \frac{l_P^2}{2}\frac{\overleftarrow\delta}{\delta
{e}_{r-}}\right)}\\
{\bar \Phi}{\widehat a}_{r-}&=&{\bar \Phi}{\left( -i\gamma
\frac{l_P^2}{2}\frac{\overleftarrow\delta}{\delta {\bar
e}_{r+}}\right)} 
\eea
for left-acting operators. However, when one
tries to find the conditions upon the inner product imposed by the reality
condition for the metric:
\be
{\langle \Phi|{\widehat e}_{r+}|\Psi\rangle}=
{\langle \Phi|{\widehat e}_{r-}|\Psi\rangle}
\ee
this leads to:
\be
e_{r+}=e_{r-}
\ee
contradicting the assumption that the functions are holomorphic, as 
initially stated.

\section{The ground state }\label{ground}
Having established the formalism we now derive the wave functions
for the ground and particle states 
of gravitons in the connection representation.
In the Bargmann representation (diagonalizing operators $G^\dagger_{rp}$)
the vacuum wave functions are just $\Psi=1$, whereas 
particle states are monomials in their respective variables~\cite{paper}. 
In this Section we will rederive these
wave functions in the holomorphic connection representation defined
in the last Section. Ordering issues affect the vacuum energy and fluctuations,
but not the form of the wave functions. The cases $\gamma\neq\pm i$ and 
$\gamma=\pm i$ are very different and will be discussed 
separately. While $\gamma\neq \pm i$ leads to straightforward Gaussian
wavefunctions, the case $\gamma=\pm i$ requires the introduction 
of a new mathematical tool.

\subsection{Wave functions for $\gamma\neq \pm i$}
In~\cite{paper} we found a set of annihilation operators 
$G_{r{\cal P}}$ in terms of metric and connection operators. 
To obtain the physical vacuum we must solve 
$G_{r{\cal P_+}} \Psi_0=0$, which translates into:
\be {\left({\widehat a}_{r+}- k(r+i\gamma){\widehat
e}_{r+}\right)}\Psi_0=0\; .\ee 
Using (\ref{newarp}) and (\ref{newerp}) this equation becomes,
in the connection representation:
\be {\left({a}_{r+}-
k(r+i\gamma)i\gamma\frac{l_P^2}{2} \frac{\delta}{\delta {\bar
a}_{r-}} \right)} \Psi_0=0\; .\ee This has solution: \be \label{gsphys}
\Psi_0(a_{r+},{\bar a}_{r-}) ={\cal N} \exp\left[{\frac{2}{i\gamma
(r+i\gamma) k l_P^2} a_{r+} {\bar a}_{r-}}\right] \ee where ${\cal N}$ is
a normalization constant (which is finite, as discussed in the next Section).

The counterpart vacuum condition for unphysical modes follows
from $G_{r{\cal P_-}}\Psi_0=0$, leading to: \be {\left({a}_{r+}-
k(r-i\gamma)i\gamma\frac{l_P^2}{2} \frac{\delta}{\delta {\bar
a}_{r-}} \right)} \Psi_0^{unph}=0\; ,\ee with solution: \be
\Psi_0^{unph}(a_{r+},{\bar a}_{r-}) ={\cal N}
\exp\left[{\frac{2}{i\gamma (r-i\gamma) k l_P^2} a_{r+} {\bar
a}_{r-}}\right]\; . \ee Unsurprisingly, the two conditions are
inconsistent, i.e. we cannot find a vacuum simultaneously for
physical and unphysical modes. Weeding out the
unphysical modes does not amount to setting them to the vacuum
state, but instead to factoring them out of the Hilbert space.
We will discuss this matter further in the next Section, where we will
also explicitly show that the unphysical wave functions are not normalizable.

Particle states can be constructed by acting
with creation operators upon the vacuum. These operators are~\cite{paper}:
\be
\label{Gr+dagger}G^{\dagger}_{r{\cal P}}=\frac{r}{i\gamma}({\widehat
a}^\dagger_{r-}- k(r-{\cal P} i\gamma){\tilde e}^\dagger_{r-})\; .\ee
Therefore we have:
\be
\Psi_n\propto {\bar a}_{r-}^n \Psi_0 \ee for physical gravitons,
and \be \Psi_n^{unph}\propto {\bar a}_{r-}^n \Psi_0^{unph} \ee for
the unphysical modes. 

\subsection{The singular case $\gamma=\pm i$}\label{groundsd}
For the SD and ASD connections we see that two of the four Gaussian wave 
functions derived above become ill-defined (the denominator in the
exponent is zero). The origin of this singularity is interesting.
We recall~\cite{prl} that the graviton operators are now:
\bea
\widehat g_{r+}(\vk)&=&{\widehat a}_{r+}(\vk)\label{op1}\\
\widehat g^\dagger_{r+}(\vk)&=&-{\widehat a}^\dagger_{r-}(\vk) +2kr {\widehat e}_{r-}^\dagger(\vk)\\
\widehat g_{r-}(\vk)&=&-{\widehat a}_{r+}(\vk) +2kr {\widehat e}_{r+}(\vk)\\
\widehat g_{r-}^\dagger(\vk)&=&{\widehat a}^\dagger_{r-}(\vk) \label{op4}\eea 
so that diagonalizing the connection entails diagonalizing
the annihilation operator $\widehat g_{r+}$. 
Note that since $a_{rp}$ and $a_{rp}^\dagger$ commute 
no variation of the prescription in Section~\ref{hol-solution} would get us
out of this conclusion. Therefore by going to the 
connection representation, we are forced to face a notorious problem:
that of defining the dual 
of the Bargmann representation, i.e. a representation diagonalizing 
annihilation instead of creation operators 
(see, e.g.~\cite{ribeiro,ribeiro1}).

\subsubsection{The dual of the Bargmann representation for a harmonic 
oscillator}
We illustrate the issues surrounding the definition of a dual
to Bargmann's representation resorting to the simple harmonic 
oscillator. Several attitudes can be adopted. In~\cite{ribeiro,ribeiro1}
it was advocated that one should abandon the concept of a dual space 
of ``bra'' vectors, from which the dual functions are derived. 
It was shown that it was possible to define a set of 
holomorphic functions, ${\bar\Psi}({\bar w})$, conjugate or dual  
to the holomorphic Bargmann functions, 
$\Psi(z)={\langle z|\Psi\rangle}$, which are {\it not}
${\langle \Psi | w \rangle}$, since $| w \rangle$ is left undefined.
With this strategy the dual functions are still  
entire functions, i.e. 
analytic over the whole complex domain. Particle states are
now inverse powers $1/w^n$ (cf. eqn (25) of~\cite{ribeiro1}) whereas 
the vacuum is a constant.  The associated inner product was
identified for suitably chosen contours of integration (cf. 
eqn (46) in~\cite{ribeiro1}). The result is somewhat cumbersome, but 
remains a distinct possibility.

An alternative is often used by the quantum gravity community,
but its radical mathematical nature is not often spelled out 
(see~\cite{ashbook}, however). The idea is to work with a space
of ``functions'' which are holomorphic in
the sense defined in the previous Section (functions of ``$z$'' but
not ``$\bar z$''), whilst dropping the requirement that they be entire
functions. In fact, as we shall see, we should not even require these
``functions'' to be distributions in the usual sense (or, at least, we
should broaden the concept of distribution~\cite{ashbook}). The matter
is closely related to the definition of a ``holomorphic delta function''.
A number of mathematical identities valid in the Bargmann 
formalism will cease to be
true. However there are also advantages with respect to the approach
of~\cite{ribeiro}.

\subsubsection{Holomorphic delta function}
Let us define the dual of the Bargmann representation by:
\be
a\Psi(w)=w\Psi(w)\; ,
\ee
in contrast with the usual $a^\dagger\Psi(z)=z\Psi$ (for which
the physical vacuum is $\Psi_0=1$, particle states $\Psi_n\propto z^n$,
and the inner product measure is $e^{-\bar z z}$). 
From $[a,a^\dagger]=1$ we have:
\be\label{adaggerdual}
a^\dagger \Psi(w)=-\frac{d\Psi(w)}{dw}\; .
\ee
Solving for the vacuum 
leads directly to a definition for the ``holomorphic delta function''.
The vacuum equation $a\Psi_0=0$ results in $w\Psi_0(w)=0$,
suggesting $\Psi_0=\delta(w)$ with the first defining property:
\be
w\delta (w)=0\; .
\ee
The second defining property may be obtained from the fact that norms
are independent of the representation. 
The inner product in the dual representation may be derived
as usual~\cite{bargrev} 
by formally requiring ${\langle \Phi|a^\dagger |\Psi\rangle}=
\overline{\langle \Psi| a |\Phi \rangle}$). However, this 
exercise leads to:
\be
{\langle\Phi|\Psi\rangle}=\int dw\, d{\bar w}\,
e^{w{\bar w}}{\bar \Phi}({\bar w})\Psi(w)\ee
which differs by the sign in the exponent with regards to the measure
for the Bargmann representation. Since:
\be
{\langle 0|0\rangle}=
\int dz\, d{\bar z}\,
e^{-z{\bar z}}=
\int dw\, d{\bar w}\,
e^{w{\bar w}}{\bar \delta}({\bar w})\delta(w)\;, \ee
we should impose the second defining property for the holomorphic
delta function:
\be\label{seconddef}
\int dw \, d{\bar w}\,  e^{w{\bar w}}
{\bar \delta}({\bar w}) \delta (w)=1 \; .\ee
This delta function is obviously an odd object. It is not analytical,
or even a distribution in the usual sense.  It only makes sense when
integrated multiplied by its complex conjugate. In a sense, it is the 
``square root of a distribution''. It should not
be confused with the ``delta'' function used in the Bargmann
formalism~\cite{bargrev}, 
which is better described as a ``Reproducing Kernel'' or
a ``Principal Vector'', and is given
by the entire function $K(z;w)= e^{\bar z w}$.

\subsubsection{Further properties of the complex delta function}
We stress again that in the dual representation just
defined, wave functions are no longer entire functions
and so many commonly used identities in the Bargmann formalism 
(e.g. those involving the Reproducing Kernel) are no longer valid.
This is to be contrasted with the approach in~\cite{ribeiro,ribeiro1}.
Instead one must become acquainted with the algebraic properties of 
the holomorphic delta function. From (\ref{adaggerdual}) we see that
the particle states are now:
\be
\Psi_n(w)={\langle w|n\rangle}=\frac{ (-1)^n}{\sqrt{n!}}\frac{d^n \delta (w)}
{dw^n}\; .
\ee
It can then be proved (either by direct integration
by parts, or using the invariance of the inner product with respect
to the representation) that:
\be
\int dw \, d{\bar w}\,  e^{w{\bar w}}
\frac{d^n \bar \delta (\bar w)}
{d\bar w^n} \frac{d^m \delta (w)}
{dw^m}= \frac{\delta_{nm}}{n!}\; .
\ee
Likewise, considering unphysical modes in the Bargmann representation
(those belonging to the ``Dirac sea'', as in~\cite{wittenym}), we
can infer that:
\be
\int dw \, d{\bar w}\,  e^{-w{\bar w}}
{\bar \delta}({\bar w}) \delta (w)=\infty \; .\ee
In a similar fashion many identities of this sort may be derived,
establishing the basic rules of calculus for the holomorphic 
delta function. 

It is also possible to write the holomorphic delta function in a more explict 
form. Consider the relationship between the Bargmann representation with eigenstates $|z\rangle$ (for which the measure is $e^{-z\bar{z}}$) and the dual we just defined. If we want to transform between the two, we can write a state $\langle w|\Psi\rangle$ in terms of $\langle z|\Psi\rangle$ as
\be
\langle w|\Psi\rangle=\int dz d \bar{z} e^{-z\bar{z}} \langle w|z\rangle\langle z|\Psi\rangle \ee
The ground state in the two representations is given by a constant and $\delta(w)$, respectively, so that:
\be\label{gstransform}
\delta(w)=\int dz d \bar{z} e^{-z\bar{z}} \langle w|z\rangle\; . \ee
The first excited state is $-\frac{d}{dw} \delta(w)$ in the dual 
representation and $z$ in the Bargmann representation, so:
\bea\label{festransform}
-\frac{d}{dw}\delta(w)&=&\int dz d \bar{z} e^{-z\bar{z}} \langle w|z\rangle z \nonumber\\
&=& -\int dz d \bar{z} e^{-z\bar{z}} \frac{d}{dw}\langle w|z\rangle \eea
where the second identity comes from differentiating (\ref{gstransform}) with respect to $w$.
Therefore we obtain a differential equation for the (non-analytic) inner product $\langle w|z\rangle$:
\be \langle w|z\rangle z=- \frac{d}{dw} \langle w|z\rangle \ee

This has solution $\langle w|z\rangle=e^{-wz}$, implying:
\be
\delta(w)=\int dz d \bar{z} e^{-z(w+\bar{z})} \; .\ee

\subsubsection{Wave functions for $\gamma=\pm i$}
We have illustrated our method with the simple harmonic
oscillator, but what we've said so far transposes directly to the 
wave functions of gravitons in the connection representation,
when $\gamma=\pm i$. For definiteness we discuss the SD ($\gamma=i$)
case (the wave functions don't change if $\gamma=-i$, but what
one calls physical and unphysical modes does change, as shall be seen in the 
next Section).  With the definition for the holomorphic delta function
just provided, the equations for the physical modes
$R+$ and $L-$:
\bea
{\widehat g}_{R+}\Psi_0&=&{\widehat a}_{R+}\Psi_0={a}_{R+}\Psi_0=0\\
{\widehat g}_{L-}\Psi_0&=&(-{\widehat a}_{L+}+ 2kr {\widehat e}_{L+})\Psi_0\nonumber\\
&=& {\left(-a_{L+} + k l_P^2 \frac{\delta}{\delta {\bar a_{L-}}}
\right)}\Psi_0= 0 \eea 
can be solved as:
\be \label{Psi0SD}\Psi_0={\cal N}\delta (a_{R+})
\exp{\left[\frac{a_{L+}{\bar a}_{L-}}{k l_P^2} \right]}\, . \ee 
The unphysical modes, on the other hand, can be obtained from:
\bea
{\widehat g}_{L+}\Psi_0&=&{\widehat a}_{L+}\Psi_0={a}_{L+}\Psi_0=0\\
{\widehat g}_{R-}\Psi_0&=&(-{\widehat a}_{R+}+ 2kr {\widehat e}_{R+})\Psi_0\nonumber\\
&=& {\left(-a_{R+} + k l_P^2 \frac{\delta}{\delta {\bar a_{R-}}}
\right)}\Psi_0= 0 \eea 
resulting in:
\be \label{Psi0SDunp}\Psi_0={\cal N}\delta (a_{L+})
\exp{\left[\frac{a_{R+}{\bar a}_{R-}}{k l_P^2} \right]}\, . \ee 
Particle states are products of exponentials and monomials,
or derivatives of the delta function, as appropriate.

\section{Selection of physical states}\label{norms}
As in the graviton representation, we need to show that the physical/unphysical states are normalizable/non-normalizable. 
This requires identifying the inner product. 
In the graviton representation, physical and unphysical graviton 
states depend on different variables ($z_{r{\cal P}_+}$ and 
$z_{r{\cal P}_-}$, respectively). Evaluating inner products as integrals,
the measure splits into a $z_{r{\cal P_+}}$-dependent part belonging 
to physical states and a $z_{r{\cal P_-}}$-dependent factor
corresponding to unphysical states. The same cannot be said for the 
connection representation, as all states depend on common variables  
$a_{r+}$ and $\bar{a}_{r-}$. The measure therefore can't be split in the 
same way. 

A generalization of the usual procedure, applicable to the connection
representation, consists of applying the torsion-free condition in the 
integration leading to the physical inner product.  
Note that in the graviton representation no torsion
implies $G_{r{\cal P_-}}\approx 0$, and therefore $z_{r{\cal P_-}}\approx 0$. 
When we factor out the $z_{r{\cal P_-}}$ integration, to claim that the
physical modes are normalizable, we are therefore
applying this prescription. In the connection representation the 
torsion-free condition implies: 
\be\label{aonshell} a_{r-}=\frac{r+i\gamma}{r-i\gamma}a_{r+} \; . \ee
The physical inner product is obtained by inserting (\ref{aonshell}) 
into (\ref{solmu}), eliminating $a_{r-}$. We can then use 
this inner product to check whether physical graviton states are normalisable.

For $\gamma \neq \pm i$  (applying Eq.~(\ref{aonshell})
to the states as well) we find for the physical vacuum 
\be {\langle \Psi_0 |
\Psi_0\rangle}=\int \prod_r d a_{r+} d {\bar a}_{r+} 
\exp\left[\sum_r -\frac{4 a_{r+}{\bar a}_{r+} }{kl_P^2(r-i\gamma)^2}
\right]\; .\ee
This converges for all values of $r$ and $\gamma$ as required. Using the same
prescription for
the unphysical ground state $\Psi_0^{unph}$ we obtain
\bea {\langle \Psi_0^{unph} |
\Psi_0^{unph}\rangle}&=& \nonumber \\ \int \prod_r d a_{r+} d {\bar a}_{r+} 
&\exp&\left[\sum_r \frac{4}{kl_P^2(r-i\gamma)^2}a_{r+}{\bar a}_{r+} \right]
\; \eea
so these states aren't normalisable for any $\gamma$, as expected. We can also compute
\be {\langle \Psi_0 | \Psi_0^{unph}\rangle} = 0 \; , \ee
implying that
physical and unphysical states are orthogonal.


For the SD connection ($\gamma = i$), the measure (\ref{solmu}) reduces to 
\be
\mu=
\int d^3k
\Big(\frac{1}{l_P^2 k}|a_{R+}|^2-\frac{1}{l_P^2 k}|a_{L-}|^2\Big)\; .\ee
In this case the torsion-free condition is 
$a_{R-}=a_{L+}=0$, so again we are factoring out of the integrals 
some of the variables. Applying the same prescription as above, the 
norm of the physical ground state (\ref{Psi0SD})
is given by:
\bea &&{\langle \Psi_0 |\Psi_0\rangle}=\nonumber\\
&&\int d a_{R+} d {\bar a}_{R+} 
\exp\left(\frac{1}{kl_P^2}|a_{R+}|^2\right)\delta (a_{R+})\bar{\delta} (\bar{a}_{R+})\nonumber \\
&& \int d a_{L-} d {\bar a}_{L-} 
\exp\left(\frac{-1}{kl_P^2}|a_{L-}|^2\right) \; .\label{normSD}\eea
The second integral obviously converges and the first integral is also 
finite due to equation (\ref{seconddef}). 
As in the more general case, the unphysical vacuum (\ref{Psi0SDunp})
is non-normalizable and orthogonal to the physical vacuum state.

\section{Why the Kodama state can never describe gravitons}
\label{kodama}
In the preceding Sections we studied the connection representation
for gravitons and their vacuum.
It was suggested in~\cite{leereview} that the perturbed 
Kodama state could represent gravitons in a de Sitter background. The claim
was further examined in~\cite{leelaur}, where evident contradictions 
began to surface. 
In this Section the conflict between the two is rendered explicit and 
explained. 
In Appendix~\ref{appendix1} we derive the perturbed Kodama state 
using the same set of conventions we have used for deriving gravitons states.
The outcome is
\be
\Psi^{KOD}={\cal N}\exp{\left(\frac{2 i\gamma}{l_P^2 H^2a^2} \sum_r (kr -\gamma Ha) a_{r+}{\bar a}_{r-} \right)} \; ,\ee
to be contrasted, in the limit $|k\eta|\rightarrow \infty$,
with the wave functions presented in Section~\ref{ground}.

The conflict is far from surprising.
An algebraic argument is given in Section~\ref{simple}  
showing that the perturbed Kodama state isn't even an 
eigenstate of the perturbative Hamiltonian, as gravitons are.
It represents self-dual excitations, combining the positive
frequency half of the right-handed graviton expansion, and the negative 
frequencies of the left graviton. 
Such self-dual states are not
physical states, as we prove explicitly in Section~\ref{Kodnorm}.

\subsection{A  simple argument}\label{simple}
There is a very simple algebraic reason why the perturbed Kodama
state cannot represent gravitons or their ground state. 
The argument is valid even before issues such as the inner product
and normalizability are brought into play. The argument concerns the
relation between the Hamiltonian constraint  and 
the self-dual operator
{\it when re-examined at the perturbative level}. 

Schematically (dropping integrals, summations, 
proportionality constants and contractions with $\epsilon$, 
irrelevant for the argument), the Hamiltonian constraint,
when $\gamma=\pm i$ and in the presence of a cosmological constant,
takes the form:
\be\label{ham} {\cal H}= EE{\cal S} \ee where the
``self dual'' operator~\footnote{The
term self-dual is often used in two senses in this context. The connection $A$
is always self-dual when $\gamma=i$. However, ${\cal
S}$ resembles a ``self-dual'' operator if $H^2=i$.}  
is \be\label{selfd} {\cal S}=B+H^2 E\; ,
\ee 
(see Eq.~44 in~\cite{paper} with $\gamma=\pm i$, for example). 
Here $B$ is the magnetic field of $A$ and we
have assumed an ``EEF'' ordering for the Hamiltonian
constraint. The Kodama (or Chern-Simons) wave function is
annihilated by ${\cal S}$ and therefore it is
a solution to the Hamiltonian constraint with this ordering.

It can be explicitly checked that this
is true to zeroth order, considering a de Sitter 
solution (e.g.~\cite{paper}, Section IIA). 
However, the relation between self-dual states and solutions to the 
Hamiltonian constraint breaks down in perturbation theory. 
We use a notation (already used in~\cite{prl,paper}) where left-side 
superscripts denote the order of a quantity and left-side subscripts the
order of the metric and
connection quantities on which it depends. Thus, the second order
Hamiltonian (${}^2 {\cal H}$; no left subscript specified)  contains
terms quadratic in first order variables (denoted ${}^2_1 {\cal H}$)
and terms linear in second order variables (${}^2_2 {\cal H}$),
that is: 
\be 
{}^2 {\cal H}={}^2_1 {\cal H} + {}^2_1 {\cal H}.
\ee
As explained in~\cite{paper},
perturbation theory is ruled by ${}^2_1 {\cal H}$. 
Gravitons and their vacuum are eigenstates of ${}^2_1 {\cal H}$. The term
${}^2_1 {\cal H}$ is called by cosmologists the ``backreaction''.
Only the full ${}^2 {\cal H}$ needs to be weakly zero; thus
gravitons have dynamics, with the Hamiltonian constraint being enforced
to second order by the backreaction.

The perturbed Kodama state ${}^2_1\Phi$ (see Appendix~\ref{3kodamas})
is annihilated by ${}^1_1
{\cal S}$. Such a state (or any eigenstate of ${}^1_1
{\cal S}$) can never be an eigenstate of ${}^2_1 {\cal H}$. 
This is obvious even without inspecting the complicated perturbative
expressions for
both. Schematically, again, 
perturbing (\ref{ham}) produces an expansion of the 
form:
\be {}^2_1{\cal
H}=2({}^0E) ({}^1_1E) ({}^1_1{\cal S}) + ({}^1_1E) ({}^1_1E)
({}^0{\cal S}) + ({}^0 E) ({}^0 E) ({}^2_1{\cal S}) \nonumber\ee and
obviously ${}^1_1{\cal S}$ does not factorize on the right,
{\it as it does non-perturbatively}. Therefore a state annihilated
by ${}^1_1{\cal S}$ is not and an eigenstate of ${}^2_1{\cal
H}$, and so the perturbed Kodama state cannot
represent gravitons or their vacuum.

Even though this is a new result, it is hardly surprising.
It was pointed out in~\cite{paper} (Section II),
reviving an old result~\cite{ash0},
that gravitons are not self-dual or anti-self-dual. Once positive
and negative frequencies are included in the expansions (an omission
behind much confusion in the literature), one finds that self-dual
states combine the positive frequency of the right handed
graviton and the negative frequency of the left-handed
anti-graviton, before reality conditions are imposed. These are
the states described by the perturbed Kodama state: a
composite of half gravitons, which can never be physical or normalizable, 
once the inner product is fixed by the reality conditions, 
as we'll prove explicitly in Section~\ref{Kodnorm}.

It is interesting to 
contrast Yang-Mills and gravity with regards to
this argument. In Yang-Mills theories the Hamiltonian, again
schematically, is of
the form: \be {\cal H}={\cal S}^\star {\cal S}\; , \ee with
self-dual operator: \be {\cal S} = E+iB\; .\ee 
Perturbatively this becomes:
\be \label{hamympert}{}^2_1{\cal H}=({}^1_1{\cal S})^\star
({}^1_1{\cal S})\; , \ee and since ${}^1_1{\cal S}$ factors on the right
one might think that, unlike with gravitons, gauge bosons are indeed
eigenstates of the perturbed self-dual operator. One might then
be tempted to argue that self-dual and
anti-self-dual states correspond to positive and negative
helicities and that the Chern-Simons state (solving ${\cal H}\approx 0$)
represents states with equal numbers of gravitons with positive and 
with negative energy, or a ground state devoid of vacuum energy.
This is essentially the argument in~\cite{wittenym}. 

It turns out that the
argument is not only inapplicable to gravity (due to the more intricate nature
of its Hamiltonian) but is also incorrect in Yang-Mills theories, due to
quantum mechanical ordering issues. The ordering leading to 
(\ref{hamympert}) is non-locally distinct from the ordering leading to
boson operators. The non-alignment of (anti-)self-dual states and graviton
states applies to gauge bosons, too, and the perturbed 
Chern-Simons state describes non-physical composites of half
gauge bosons\footnote{We thank Abhay Ashtekar for pointing this out to us.}.

\subsection{Non-normalizability of the Kodama state fluctuations}\label{Kodnorm}

Here we explicitly prove that the Kodama state,
\be
\Psi^{KOD}={\cal N}\exp{\left(\frac{2 i\gamma}{l_P^2 H^2a^2} \sum_r (kr -\gamma Ha) a_{r+}{\bar a}_{r-} \right)} \; ,\ee
 is non-normalisable, adopting the inside horizon limit ($k\gg Ha$). As it is only defined for $\gamma=\pm i$, only two components of the connection remain once the torsion-free condition is invoked. For $\gamma=i$, these are $\widehat{a}_{R+}$ and $\widehat{a}_{L-}$ with $\widehat{a}_{R-}=\widehat{a}_{L+}\approx0$. Therefore, when computing the inner product of the Kodama state on-shell most of the terms vanish and we are left with
\bea \langle \Psi^{KOD} | \Psi^{KOD} \rangle = &&\int da_{R+} d{\bar a}_{R+}
 \exp \left(\frac{|a_{R+}|^2}{kl_P^2} \right) \nonumber \\
&&\int da_{L-} d{\bar a}_{L-}
 \exp \left(-\frac{|a_{L-}|^2}{kl_P^2} \right)\; .\eea
While the second term converges, the first term clearly diverges and the state is not normalisable. A similiar result is obtained for $\gamma=-i$. Hence it can be seen that the perturbed Kodama state is not normalisable under the inner product previously identified, and our criterium implies that the excitations identified in the previous subsection  belong to the unphysical Hilbert sub-space.

\section{Concluding remarks}\label{specs}
In this paper we have derived a perturbative quantum ground state for tensor 
fluctuations in de Sitter space-time. We did so in the same set up leading
to the Kodama state but also within the more general framework of any 
representation diagonalizing the connection. We identified perturbative wave
functions for all values of $\gamma$ and factor orderings (factor 
ordering changes the vacuum energy and fluctuations but not the
form of the wave functions; see~\cite{paper}).
We also identified the inner product, and the physical and non-physical
modes. Two things are immediately evident about what these states are {\it not}.
They're {\it not} the Bunch-Davis vacuum: our vacua depend on $\gamma$ and 
for each value display a distinct chirality, reflected in the vacuum energy and
fluctuations, in contrast with the Bunch-Davis vacuum. They're also {\it not}
the perturbed Kodama state, as can be seen by direct inspection. 
Our states represent new solutions which, we argue, are
the physically correct ones for discussing phenomenology in the 
Ashtekar formalism, at least in this representation.

Even though rejecting the Kodama state has by now become pass\'e,
our rejection has a strong object lesson to it, which is why we
devoted Section~\ref{kodama} to
contrasting our results and the perturbed Kodama state.
Reality is often built from perturbation theory in a ``Russian
doll'' set up where quantities order by order appear non-trivially
in non-linear expressions (such as the Hamiltonian constraint) expanded
to a certain order. For example, in~\cite{paper} it was argued that 
the infamous ``problem of time'' in quantum
gravity naturally fades away if one adopts a more practical perspective
based on perturbation theory. The second order
Hamiltonian is made up of terms quadratic in first order variables (to
be used in perturbation theory) and terms linear in second order variables
(the ``backreaction'', to be ignored). The Hamiltonian constraint 
applies to the full second order Hamiltonian (including the backreaction), 
not to the Hamiltonian relevant for
perturbation theory. Pragmatically, and somewhat ideosyncratically, 
it can be claimed that the problem of time is an illusion of non-perturbative 
theory, to be ignored in practice. 

A similar sleight of hand in perturbation theory sees
the self-dual operator factor non-perturbatively, but not perturbatively,
in the Hamiltonian constraint. For this reason,
the perturbed Kodama state is not a solution to the perturbative
Hamiltonian. This convincingly disproves the Kodama state as the correct 
path to phenomenology in quantum gravity. The perturbed Kodama state cannot
describe gravitons and its excitations are unphysical states.
A theory of quantum gravity without gravitons is not
a theory of quantum gravity at all.
The implications are deep. The Kodama state is a semi-classical solution
to the theory. Therefore
it appears that (at least in this case) {\it the semi-classical limit is 
the wrong path to reality, which may or may not be perturbative, but which 
is nonetheless fully quantum}. This may well happen more generally, 
raising alarm bells about using WKB solutions to make contact with reality
in quantum gravity theories. If nothing else, a
 moral lesson may be drawn from this paper:
using the semi-classical approximation as a bridge to phenomenology
may be tremendously unprofitable.

Obviously much work remains to be done. Somewhat trivially one may ask whether
our construction could be extended to scalar and vector linear fluctuations, 
and with what implications for cosmology. More intriguing is the possibility 
that what we have uncovered in this paper is merely the tip of a 
non-perturbative iceberg, hinting at a new ground state in the {\it full} 
quantum gravity theory. 
We may conjecture that our perturbative solution is a linear
approximation to 
a full non-perturbative solution. Whilst we have been unable
to derive this state, we can infer some properties about it.
It should be chiral. It should violate CPT, at least face value. 
The possibility remains that one might prove a no-go theorem regarding
the existence of this non-perturbative 
state. Until that happens, it is an interesting
challenge to work out its expression, as well as its associated inner product.

\begin{acknowledgements}
We'd like to thank Abhay Ashtekar, Carl Bender, Laurent Freidel, Chris Isham, 
Carlo Rovelli and Lee Smolin for discussions regarding this project.   
\end{acknowledgements}

\appendix 

\section{The perturbed Kodama state in Fourier space}
\label{appendix1}
In this Appendix we derive an explicit expression for the perturbed
Kodama state in Fourier space. We aim to do so with the same set of
conventions we used for deriving graviton modes, and so start by reviewing
these briefly (referring the reader to~\cite{paper} for more details). 
We recall that the point of 
our expansions was the introduction of a priori independent positive and
negative frequencies, and the introduction of boundary conditions ensuring
that $\vk$ is indeed the direction of motion of a mode (rather than $-\vk$).
The former point is essential in identifying all the modes in a theory which
starts off complex. The latter point ensures that the 
correct physical polarizations
are assigned to each
mode (R and L don't mean anything until we know the direction
of motion); it also removes spurious pump terms inside the horizon
(particle pair production).

\subsection{Hamilton's equations in Fourier space}
In writing the Kodama state in Fourier space there is an 
an issue affecting the sympletic structure and Hamilton's equations
to that cannot be ignored when writing the perturbed self-dual equations
in Fourier space. When transitioning from
position space to Fourier space, we use expansions~\cite{paper}:
\bea \delta e_{ij}&=&\int
\frac{d^3 k}{(2\pi)^{\frac{3}{2}}} \sum_{r}
\epsilon^r_{ij}({\mathbf k}) {\tilde\Psi}_e(\vk,\eta)e_{r+}(\vk)
\nonumber\\
&&
+\epsilon^{r\star}_{ij}({\mathbf k}) {\tilde\Psi}_e^\star (\vk,\eta)
e^{\dagger}_{r-}(\vk)\nonumber\\
a_{ij}&=& \int \frac{d^3 k}{(2\pi)^{\frac{3}{2}}} \sum_{r}
\epsilon^r_{ij}({\mathbf k}) {\tilde\Psi}_{a}^{r+}(\vk,\eta)a_{r+}(\vk)
\nonumber\\
&& +\epsilon^{r\star}_{ij}({\mathbf k}) {\tilde\Psi}_{a}^{r-
\star} (\vk,\eta)a^{\dagger}_{r-}(\vk)\; . \label{fourrier}\eea
The virtues of these expansions were extolled in~\cite{paper}. 
They avoid the embarrassment of finding that the reality
conditions constrain the number of gravitons 
moving in opposite directions~\cite{leelaur}. 
The central point in this respect is
the use of independent positive and negative frequencies---i.e. ensuring
that the theory initially contains gravitons and anti-gravitons.
The amplitudes carry two indices: $r$ for helicity and
$p$ for graviton/anti-graviton. All the reality conditions then do
is to identify gravitons and anti-gravitons, mode by mode: 
$\vk$ by $\vk$, helicity by helicity. 

It is also crucial that we require that $\vk$
label the direction of motion (in the sense that $-\vk$ should
label the {\it opposite} direction of motion). This amounts to boundary
condition:
\be\label{ketabig}\Psi(k,\eta)\sim{e^{-i k\eta}}\ee when
$|k\eta|\gg 1$ for both $+\vk$ and $-\vk$ directions
(where ${\tilde \Psi}(\vk,\eta)=\Psi(k,\eta) e^{i\vk\cdot \vx} $).
However, when we insert expansions (\ref{fourrier}) 
into the Hamiltonian, it knows
nothing about the boundary condition, so quite naturally its Hamilton's
equations try to spit
out two types of modes for each $r,p$: 
$\Psi(k,\eta)\sim{e^{\pm i k\eta}}$. And obviously there is then
a coupling between $\vk$ and $-\vk$ modes, because some of them
are the same physical modes, only written down differently. Therefore we
have to modify the Hamiltonian in Fourier space in order to account for
the boundary condition implicit in (\ref{fourrier}).

To illustrate the issue more explicitly, 
let's take Hamilton's equations in position space
for $\gamma=\pm i$:
\bea a'_{ij}&=&2
\gamma H^2 a^2\delta e_{ij}-\gamma \epsilon_{inm}
\partial_n a_{mj}\label{ham1}\\
\delta e'_{ij}&=&-\gamma(a_{ij}-\epsilon_{inm}\partial_n\delta
e_{mj})\; . \label{ham2}\eea
If we Fourier transform them according to (\ref{fourrier}),  
assuming that all the modes are independent and satisfy the right
boundary conditions, we obtain:
\bea
{\tilde a}_{rp}'(\vk)&=&\gamma p (-rk{\tilde a}_{rp}(\vk)+ 2H^2a^2
{\tilde e}_{rp}(\vk))\\
{\tilde e}_{rp}'(\vk)&=& -\gamma ({\tilde a}_{rp}(\vk)-rpk {\tilde e}_{rp}
(\vk)) \; .
\eea
Then, so that boundary condition (\ref{ketabig}) is met, we must 
have $a_{rp}=0$ when $i\gamma pr = 1$. We also recover the results:
\bea
\Psi_e'' + (k^2 -2H^2a^2)\Psi_e&=&0\\
\Psi^a_{rp}&=&\gamma p\Psi_e' +rk \Psi _e
\eea

However, if we insert
expansions (\ref{fourrier})  into  the Hamiltonian (Eq.~17 of~\cite{prl}) 
we'll find for modes inside the horizon:
\bea\label{hameff} {\cal
H}_{eff}&=&\frac{1}{l_P^2}\int d^3k\sum_r
g_{r-}(\vk)g_{r+}(-\vk)+g_{r-}(\vk)g_{r-}^\dagger(\vk)\nonumber\\
&+&g_{r+}^\dagger(\vk)g_{r+}(\vk)+
g_{r+}^\dagger(\vk)g_{r-}^\dagger(-\vk)\; , \eea
with graviton operators defined as in (\ref{op1})-(\ref{op4}). With commutators 
(\ref{fixedcrs}) we therefore obtain equations:
\bea
{\dot a_{r+}}(\vk)&=&-\gamma rk(a_{r+}(\vk)+a^\dagger_{r-}(-\vk))\\
{\dot a^\dagger_{r-}}(\vk)&=&-\gamma rk(a^\dagger_{r-}(\vk)+
a_{r+}(-\vk))
\eea
that is, we get spurious couplings between the $\vk$ and $-\vk$ modes,
which reflect the fact that nowhere in the formalism have we expressed the
requirement that modes labeled by $\vk$ are moving along $\vk$,
not $-\vk$. This can be corrected by adopting the improved Hamiltonian:
\be\label{hameffimp} {\cal
H}_{eff}=\frac{1}{l_P^2}\int d^3k\sum_r
g_{r-}(\vk)g_{r-}^\dagger(\vk)
+g_{r+}^\dagger(\vk)g_{r+}(\vk) \nonumber\ee
which has been ``told'' the correct boundary condition.

\subsection{The perturbed and the full Kodama state}\label{3kodamas}
With these conventions in mind we 
now write down the Kodama state non-perturbatively and
perturbatively, first in position space then in Fourier space.
A number of issues found with Hamilton's equations are
re-encountered here, and can be resolved in the same way.

The Kodama state is the solution to the self-dual equation ${\cal S}\Phi=0$
in the connection representation, that is with:
\bea
{\hat A^i_a}(\vx)\Phi(A^i_a)&=&A^i_a(\vx)\Phi(A^i_a)\\
{\hat E^b_j}(\vx)\Phi(A^i_a)&=&-i\gamma l_P^2\frac{\delta}{\delta A^j_b(\vx)}
\Phi(A^i_a)
\eea
representing algebra:
\be\label{unfixedcrs} \left[{\hat A^i_a}(\vx),{\hat E^b_j}(\vy)\right] = i\gamma
l_P^2\delta^b_a\delta^i_j\delta(\vx-\vy)\; . \ee
It's easy to prove that the self-dual equation is satisfied by
the Kodama, or Chern-Simons state:
\be
\Phi={\cal N}\exp{\left(\frac{i\gamma}{2l_P^2 H^2}S_{CS} \right)}
\ee
with
\bea
S_{CS}&=&\int {\rm Tr} (A\wedge dA + \frac{2}{3} A\wedge A\wedge A)\nonumber\\
&&\int d^3 x \epsilon^{abc}(A^i_a \partial_b A^i_c
+\frac{1}{3} \epsilon ^{ijk} A^i_a A^j_b A^k_c)\; .
\eea
This is because it can be easily checked that:
\be
{\widehat E}^a_k S_{CS}= -2i\gamma l_P^2 B^a_k
\ee
so that:
\be
({\widehat B}^{ka} + H^2 {\widehat E}^{ka})\Phi=0
\ee
which is the self-dual equation.

These expressions
result in equivalent ones for the perturbations in real space
(see~\cite{paper} for definitions). Specifically we have:
\be
{}^2_1S_{CS}=\int \frac{d^3 x}{a^2}
(\epsilon_{ijk}a_{ni}\partial_j a_{kn}
-\gamma Ha a_{ij} a_{ij})
\ee
and since now the algebra is:
\be\label{unfixedcrs1} \left[a^i_a(\vx),\delta
e^b_j(\vy)\right] = -i\gamma
l_P^2\delta^b_a\delta^i_j\delta(\vx-\vy)\;  \ee
we should have
\bea
{\widehat a_{ij}}(\vx)\Phi(a_{ij})&=&a_{ij}(\vx)\Phi(a_{ij})\\
{\widehat {\delta e}_{ij}}(\vx)\Phi(a_{nm})&=&i\gamma l_P^2\frac{\delta}{\delta
a_{ij}(\vx)} \Phi(a_{nm})\; .
\eea
Therefore:
\be
a H^2\widehat {\delta e}_{ij}\Phi=\Phi(\epsilon_{inm}\partial_n a_{mj} -\gamma
Ha a_{ij})
\ee
and we still satisfy
\be
(\delta B_{ij}-H^2 a{\widehat \delta e}_{ij}){}^2_1\Phi=0
\ee
(which is the perturbed SD equation), because:
\be
\delta B_{ij}=\frac{1}{a}(\epsilon_{inm}\partial_n a_{mj}-\gamma Ha
a_{ij})\; .
\ee

If now we try to transpose this to Fourier modes we find a problem
similar to that described in the previous sub-section. 
If we expand the Chern-Simons action naively, we obtain for the 
second order terms quadratic in first order variables:
\bea
{}^2_1S_{CS}&=&\frac{1}{a^2}\int d^3k\sum_r 2(kr -\gamma Ha)
[a_{r+}(\vk) a_{r+}(-\vk)\nonumber\\
&&+ 2 a_{r+}(\vk){\bar a}_{r-}(\vk)
+{\bar a}_{r-}(\vk){\bar a}_{r-}(-\vk)]\; .
\eea
We find that this represents
\be
\delta B_{ij}=\frac{i\gamma a}{2l_P^2}{\widehat \delta e}_{ij}
{}^2_1 S_{CS}
\ee
but not mode by mode, independently. Indeed
\be
B_{rp}=(rk-\gamma pHa) a_{rp}
\ee
and in order to get the required
\be
B_{rp}=\frac{a^2 p}{4}\frac{\delta}{\delta {\bar a}_{r{\bar p}}}
{}^2_1 S_{CS}
\ee
we should discard the terms coupling $\vk$ to $-\vk$.
Therefore the perturbed Kodama state in Fourier space is:
\be
{}^2_1\Phi={\cal N}\exp{\left(\frac{2 i\gamma}{l_P^2 H^2a^2}
\int d^3k\sum_r (kr -\gamma Ha) a_{r+}(\vk){\bar a}_{r-}(\vk)
\right)}
\ee

\end{document}